# The Highly Disordered Zintl Phase Ca$_{10}$GdCdSb$_9$ – New Example of a p-type Semiconductor with Remarkable Thermoelectric Properties

*Michael O. Ogunbunmi and Svilen Bobev*[*]

Department of Chemistry and Biochemistry, University of Delaware, Newark, Delaware, 19716, United States

[*]Corresponding author; Email: bobev@udel.edu

**Abstract**

Ca$_{10}$GdCdSb$_9$ is a new Zintl phase with a large unit cell volume (~2500 Å$^3$) and a very complex, disordered structure, which can drive the realization of ultralow thermal conductivity in this material. The measured Seebeck coefficient, $\alpha$, for single-crystalline Ca$_{10}$GdCdSb$_9$ approaches 350 µV/K at 600 K. The experimentally determined electrical resistivity of Ca$_{10}$GdCdSb$_9$ is very low, leading to a remarkably high-power factor *PF* of 23.2 µW/cm·K$^2$ at 460 K. The extraordinary *PF* value in this material, higher than those of the currently known state-of-the-art materials within the same temperature range, suggests that the Ca$_{10}$*RE*CdSb$_9$ material system (*RE* = rare earth metal) could serve as a viable playground to harnessing new efficient thermoelectric generators.

**Introduction**

The onerous task of mitigating the devastating effects of global warming on our planet lies in the ability to significantly cut down the amount of greenhouse gasses emitted from burning fossil fuels and channel a pathway to cleaner energy sources. In this wise, thermoelectric materials that provide for the conversion of waste energy in the form of heat back to reusable energy find relevance. Such a material does not only provide a sustainable source of clean energy but also helps to meet up with the ever-increasing global energy demands. In recent years, several areas of applications of thermoelectric materials such as automobiles, electronic devices, deep space probes, etc., are already being explored. Owing to the useful technological applications of these materials, many research endeavors are thus now focused on identifying suitable candidate materials and developing several techniques and tools to optimize the performance of the few known ones[1–3].

In our day-to-day life, expressions such as *disorderly conduct* and *complex behavior* are often considered negative and generally disadvantageous. The human mind is generally attuned to a good and smooth life, while the converse is never desirable. However, it turns out that certain physical and chemical systems ride on the prevalence of these seemingly undesirable conditions to give rise to novel properties of interest[4]. Such is the scenario in frustrated magnetic systems[5,6], where the frustration of the magnetic spins produces competing interactions that cannot be simultaneously satisfied and can thus birth exotic states of matter at sufficiently low temperatures. Similarly, the presence of intricate atomic bonding and highly complex structural motifs have been found to be beneficial in achieving enhanced thermoelectric performance in candidate materials[7–9]. Here, the complex nature of the crystal structure acts as effective phonon scattering centers, thus leading to very low thermal conductivity, which in combination with high Seebeck coefficient and electrical conductivity can achieve an optimal thermoelectric figure of merit, $zT = \alpha^2\sigma/(\kappa_e + \kappa_l)$ at a given operating temperature $T$, where $\alpha$, $\sigma$, $\kappa_e$ and $\kappa_l$ are the Seebeck coefficient, electrical conductivity, electronic thermal conductivity, and lattice thermal conductivity, respectively.

Zintl phases that combine the features of narrow bandgap, complex atomic bonding, and highly disordered atomic sites are promising candidates to explore for new thermoelectric materials[7]. As demonstrated in several earlier reports[10,11], Zintl phases present the relevant electronic and transport environments that favor excellent thermoelectric performance. In the recent past, some members of the $Ca_{11}InSb_9$ structural series ("11–1–9" compounds)[12], such as $Yb_{11}TrSb_9$ ($Tr$ = Ga, In)[13–15] have been reported and studied for possible thermoelectric applications. These phases feature low $\kappa_l$ ~ 0.6 W/m·K near 1000 K, as well as a small magnitude of $\alpha$. The small magnitude of $\alpha$ here is thought to be associated with the disadvantageous compensated carrier-type present in both materials. Although the low $\kappa_l$ observed is most desirable, appropriate doping that will favor a particular dominant carrier concentration is necessary to enhance both phases' thermoelectric properties.

Subsequently, Wang et al.[16] reported on the $Ca_{10}LaCdSb_9$ and $Yb_{10}LaCdSb_9$, thereafter referred to as the "10–1–1–9" phases, being structural variants of the "11–1–9" family of compounds. The "10–1–1–9" phases crystallize in the centrosymmetric space group of *Ibam* (No. 72), different from the non-centrosymmetric space group of *Iba*2 (No. 45) of the typical compounds with the $Ca_{11}InSb_9$ structure type[12]. The initial study on $Ca_{10}LaCdSb_9$ and $Yb_{10}LaCdSb_9$ communicates an idea of doping studies that may be beneficial for tuning the nature of the dominant charge carriers and concentrations. Such studies also come with an additional thermoelectric flavor of further reduction in $\kappa_l$. True to this fact, an ultra-low $\kappa_l$ ~ 0.29 W/m·K is reported in $Yb_{10}LaCdSb_9$ at 875 K. Albeit, similar to the $Yb_{11}GaSb_9$ and $Yb_{11}InSb_9$ phases, the "10–1–1–9" materials reported by Wang et al.[16] also manifest a small magnitude of $\alpha$, thereby placing a constraint on the magnitude of their $zT$. Keynotes from these studies are the available window of opportunity to achieve a low $\kappa_l$ as well as their high-temperature stability, which are desired for high-temperature thermoelectric generators. A recent report on related "10–1–9" materials with the simplified formulae $Ca_{10}MSb_9$ ($M$ = Mn, Zn, Ga, In)[9] have shown heavy structural disorder in them, and a concomitant feature of ultralow $\kappa_l$ and enhanced Seebeck coefficient. These works further demonstrate the role of disorder in achieving much-improved thermoelectric $zT$.

Here, we push the experimental boundaries forward and present new results on efforts geared towards enhancing thermoelectric performance in the "10–1–1–9" phases. We report the thermoelectric properties of the $Ca_{10}GdCdSb_9$ and $Ca_{10}LaCdSb_9$ Zintl phases. While the crystal structure of $Ca_{10}LaCdSb_9$ is known[16], that of $Ca_{10}GdCdSb_9$ is presented for the first time. We demonstrate the promising thermoelectric properties resident in these phases and their potentials for use as efficient thermoelectric generators.

**Experimental**

**Synthesis**

The elements were used as purchased from Sigma-Aldrich and Alfa-Aesar with typical purity of ≥ 99.9 wt.%. Single crystals of $Ca_{10}GdCdSb_9$ and $Ca_{10}LaCdSb_9$ were synthesized serendipitously via a Sn-flux reaction designed to synthesize the yet unknown $Ca_{9-x}RE_xCd_4Sb_9$ ($RE$ = rare earth metal; $x \approx 1$) phases, which were conceived as analogs of the previously published $Ca_{9-x}RE_xMn_4Sb_9$ ($RE$ = rare earth metal; $x \approx 1$)[17]. The synthesis procedure involves weighing the elements Ca:Gd:Cd:Sb:Sn in the 8:1:4:9:23 ratio, respectively, inside an Ar-filled glovebox. The elemental mixtures were loaded into alumina crucibles and encased in a fused silica ampoules packed between two balls of quartz wool. The ampoules were subsequently evacuated and flame-sealed under a vacuum. The ampoules were then transferred into muffle furnaces (while kept in an upright manner), heated first to 773 K at the rate of 100 K h$^{-1}$ and kept for 6 h. This was followed by heating to 1223 K and homogenization at this temperature for 96 h. Then, slowly lowered the temperature to 873 K at the rate of 5 K h$^{-1}$ after which the fused-silica tubes were quickly removed from the furnace, flipped, and spun in a centrifuge at high speed to remove excess molten Sn. The tubes were then taken to the Ar-filled glove box to extract the crystals. The products of these reactions were typically multiphase; in the case of Gd, the mixture of several phases included $Ca_2CdSb_2$[18], $Ca_{14-x}Gd_xCdSb_{11}$ ($x \approx 1$)[19] and the title phase $Ca_{10}GdCdSb_9$. The crystals obtained from such reaction were only of micron sizes and therefore only suitable for structural elucidation based on single crystal data.

Further attempts aimed at synthesizing the original targeted phase of $Ca_{9-x}RE_xCd_4Sb_9$ ($RE$ = rare earth metal; $x \approx 1$) involved loading starting elements Ca:$RE$:Cd:Sb:Sn as during previous experiments following the elemental 8:1:12:9:70 ratio (notice the excess Cd). These reactions too did not produce the target phases. The only reactions where this approach succeeded, so far, were with $RE$ = La (5–7 mm single crystals of $Ca_{10}LaCdSb_9$), and $RE$ = Gd (~1 mm single crystals of $Ca_{10}GdCdSb_9$). We note that a slight modification to the reaction profile which involves a slower cooling rate of 2 K h$^{-1}$ from 1223 K to 873 K yield larger crystals (2–3 mm) of $Ca_{10}GdCdSb_9$. All reported structural work and property measurements presented here were conducted on such flux grown crystals. Experiments under different metal fluxes, as well as attempts to make $Ca_{10}RECdSb_9$ by direct fusion of the respective elements were also done, but it is noted these efforts were not successful to date, and that future results from the ongoing work will be published elsewhere.

**Powder X-Ray diffraction**

Powder X-ray diffraction (PXRD) measurements were conducted at room temperature on a Rigaku Miniflex diffractometer (filtered Cu K$\alpha$ radiation, $\lambda$ = 1.5418 Å), operated inside a nitrogen-filled glovebox. Small portions of the obtained single crystals were ground inside an argon-filled glove box using agate mortars and pestles. Data were collected between 5 and 75° in 2$\theta$ with a step size of 0.05° and 2 s per step counting time. PXRD measurements before and after exposure to air indicate that that both phases are stable in air over a period of up to 3 weeks.

The PXRD patterns for $Ca_{10}LaCdSb_9$ and $Ca_{10}GdCdSb_9$ are presented in Figure S1 in supporting information. The powder diffraction patterns for the samples can be matched with the theoretically generated patterns based on the crystal structures elucidated from single-crystal X-ray diffraction methods.

**Single-crystal X-Ray diffraction**

Single-crystal X-ray diffraction (SCXRD) studies were performed on a Bruker Photon III diffractometer equipped with a Mo K$\alpha$ radiation ($\lambda$ = 0.71073 Å) source. Single crystals prepared by flux reactions were immersed in Paratone-N oil and cut into suitable dimensions. The selected single crystal was placed onto a low-background plastic loop, quickly moved to the goniometer maintained at 200 K for the whole period of data collection. Measurements were carried out in batch runs with a frame width of 0.8° in $\omega$ with data collections using a Bruker software. SADABS software was used for multi-scan absorption correction. Structure solution and refinements were carried out using ShelXT[20] and ShelXL[21] (integrated into Olex2[22] graphical user interface), respectively. STRUTURE TIDY[23] software was used to standardize the atomic coordinates. Table 1 gives selected crystallographic parameters. The fractional coordinates of the atoms for $Ca_{10}GdCdSb_9$, are provided in Tables S1; crystallographic parameters for $Ca_{10}LaCdSb_9$, which mirror closely those from the earlier investigation,[16] are given in Tables S2 and S3 respectively. Selected interatomic distances for both phases are shown in Tables S4 and S5 respectively.

CCDC 2162688 and 2162689 contain the full supplementary crystallographic data for the compounds discussed in this paper. CIF files can be obtained free of charge via www.ccdc.cam.ac.uk/data_request/cif, or by emailing data_request@ccdc.cam.ac.uk, or by contacting The Cambridge Crystallographic Data Centre, 12 Union Road, Cambridge CB2 1EZ, U.K., fax +44 1223 336033.

**Elemental analysis**

Analyses were conducted to verify the elemental composition by means of X-ray energy dispersive (EDX) spectrometry. Selected single crystals were mounted on a carbon tape glued to an aluminum holder. The analyses were performed on Auriga 60 Cross Beam Scanning Electron Microscope equipped with Oxford Synergy X-MAX80 electron backscattering diffraction EDX detector. The beam current was 10 μA at 20 kV accelerating potential. Data were collected over several spots. See Figure S1 for the EDS results.

**Transport properties measurements**

Seebeck coefficient measurements were carried out using the integral method, and a constantan wire used as a reference on an SB-100 module MMR Tech. instrument. Appropriate geometry of the sample was mounted on the platform using a high-purity silver conductive paint and interfaced with the probe using the same silver conductive paint. Data were collected for temperatures between 300 and 600 K under high vacuum, with the sample exposed to a limited amount of air throughout the whole process.

High-temperature resistivity was measured within a temperature window of 300–460 K using a H-50 module sourced from MMR Tech. Measurements were made through the four-probe Van der Pauw method on samples utilizing four platinum wires and high-purity silver paint to make contacts. The same temperature variable chamber was used, and a vacuum environment was maintained throughout the experiment.

We also ensured that the various measurements were carried out a couple of times and on different crystals to confirm the reproducibility of the results.

**Thermal analysis**

Thermogravimetric/differential scanning calorimetry (TG/DSC) experiments were carried out using a TA Instruments SDT Q600. Analysis was conducted on ca. 10–20 mg single-crystalline sample of $Ca_{10}LaCdSb_9$ and $Ca_{10}GdCdSb_9$, which were loaded into high purity alumina pans and capped with lids. An inert atmosphere was maintained throughout the measurement by a constant flow (100 mL/min) of high-purity argon (Grade 5.0) gas. Following a brief equilibration period at 373 K, the samples were heated to 850 K at the rate of 200 K h$^{-1}$. The TG/DSC results are presented in Figure S2. These results show lack of any thermal events up to a temperature of 850 K, indicating the samples' stability.

**Results and discussions**

**Crystal structure**

The "10–1–1–9" structure is a derivative of the known "11–1–9" $Ca_{11}InSb_9$ type structure.[12] At the outset, it should be mentioned that although related to the structure of $Ca_{11}InSb_9$ (space group *Iba*2; No. 45), $Ca_{10}GdCdSb_9$ crystallizes with the centrosymmetric orthorhombic space group *Ibam* (No. 72) and the atoms are in an overall different arrangement. The same crystallographic setup has been reported for $Yb_{10}LaCdSb_9$ and $Ca_{10}LaCdSb_9$ [16]. Another important observation with regard to the crystal structure, which ought to be specifically pointed out here, is the fact that $Ca_{10}GdCdSb_9$ and $Ca_{10}LaCdSb_9$, although formally isostructural, feature slightly different crystallographic disorder; the chemical bonding in $Ca_{10}GdCdSb_9$ is even more intricate.

The average $Ca_{10}GdCdSb_9$ structure features 13 unique atomic positions—two Ca sites, three mixed Ca/Gd sites, two Cd sites, and six Sb sites as shown in Table 2. Not all of them are fully occupied as some sites are introduced to model the disorder. The polyanionic sub-lattice, as emphasized in Figure 1(a) contains $[CdSb_4]$ tetrahedra, diborane-like $[Cd_2Sb_6]$ fragments made up of edge-shared $[CdSb_4]$ tetrahedra, $[Sb_2]$-dimers, and of course, isolated $Sb^{3-}$ anions. The most significant difference between the $Ca_{10}GdCdSb_9$ and $Ca_{10}LaCdSb_9$ structures is the presence of positional disorder involving the terminal Sb atoms from the $[Cd_2Sb_6]$ fragments (labeled as Sb1 and Sb2) in the former. Sb1 and Sb2 atoms, which are both occupying 16$k$ Wyckoff positions are mutually "exclusive", meaning that a $CdSb_4$ tetrahedron is formed by either two Sb4 and two Sb1 or by two Sb4 and two Sb2 atoms, as shown in Figure 1(b). The additional disorder in $Ca_{10}GdCdSb_9$ may be contributed from the larger size-mismatch between $Ca^{2+}$ and $Gd^{3+}$ (ionic radii 1.12 Å and 1.053 Å, respectively)[24] as compared to $Ca^{2+}$ and $La^{3+}$ that are very close in size (ionic radii 1.12 Å and 1.16 Å, respectively), and in which case, the polyanionic sub-lattice does not need to compensate for the occupational cation disorder. Figure 1(c) presents the anionic substructure observed in $Ca_{10}LaCdSb_9$ as compared to that of $Ca_{10}GdCdSb_9$.

The Ca atoms are octahedrally coordinated by Sb atoms, while the three sites co-occupied by Ca/Gd atoms exhibit coordination by six, seven, and eight Sb atoms, respectively, as presented in Figure S3. We therefore note that the observed heavy disorder in $Ca_{10}GdCdSb_9$ will prove to be invaluable in phonon scattering and with an attendant effect of very low and thermal conductivity. In Table S3 selected interatomic distanced are presented.

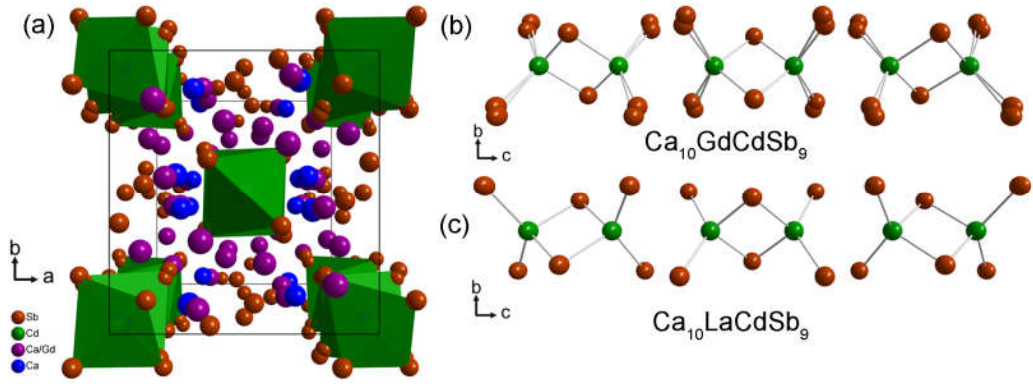

Figure 1. (a) Crystal structure of disordered $Ca_{10}GdCdSb_9$ viewed down the *c*-axis showing the Ca/Ga as well as Sb1 and Sb2 sites disorders. (b) The $[Cd_2Sb_6]^{14-}$ anions in $Ca_{10}GdCdSb_9$. Selected bond distances are presented in Table S3. (b) The $[Cd_2Sb_6]^{14-}$ anions found in $Ca_{10}LaCdSb_9$ with disorder free Sb. Some atoms such as Ca and Ca/Gd have been omitted for clarity.

**Table 1:** Selected crystallographic data of $Ca_{10}GdCdSb_9$ measured at 200(2) K with Mo Kα radiation $\lambda = 0.71073$ Å.

| | |
|---|---|
| refined formula | $Ca_{10.43(1)}Cd_{1.25(1)}Gd_{0.59}Sb_{9.01(1)}$ |
| formula weight | 1747.57 |
| crystal system | orthorhombic |
| space group | *Ibam* (No. 72), $Z = 4$ |
| $a$/Å | 11.8281(5) |
| $b$/Å | 12.4315(6) |
| $c$/Å | 16.7562(8) |
| $V$/Å$^3$ | 2463.8(2) |
| $\rho_{calc}$/ g cm$^{-3}$ | 4.71 |
| $\mu$/cm$^{-1}$ | 144.7 |
| $R_1$ ($I \geq 2\sigma_I$)$^a$ | 0.0341 |
| $wR2$ ($I \geq 2\sigma_I$)$^a$ | 0.0565 |
| $R_1$ (all data) | 0.0417 |
| $wR_2$ (all data) | 0.0588 |
| Largest difference peak and hole | 2.59; −2.16 |

$R_1=\sum||F_o|-|F_c||/\sum|F_o|$; $wR_2=[\sum[w(F_o^2-F_c^2)^2]/\sum[w(F_o^2)^2]]^{1/2}$, where $w=1/[\sigma^2 F_o^2+(0.0101P)^2+(57.9647P)]$, and $P=(F_o^2+2F_c^2)/3$.

**Transport Properties**

The temperature dependence of Seebeck coefficient $\alpha(T)$ of the two phases investigated in the temperature window of 300 to 600 K is presented in Figure 2(a). $\alpha(T)$ of both phases is positive within the temperature range studied, indicating of a p-type material behavior with dominant hole carriers. $\alpha(T)$ of $Ca_{10}LaCdSb_9$ at room temperature is only about 16.7 µV/K which is nearly 50% lower than the reported value in $Yb_{10}LaCdSb_9$[16] around the same temperature. However, $\alpha(T)$ of $Ca_{10}LaCdSb_9$ shows a strong temperature dependence upon heating with an increase in magnitude of a significant factor of ca. 17 between 300 and 600 K, with the value at 600 K being 286.2 µV/K. This value is far higher than even what is recorded for $Yb_{10}LaCdSb_9$ at 1023 K and also superior to those reported in the typical "11–1–9" phases[13] and several others with excellent thermoelectric properties such as $Yb_{21}Mn_4Sb_{18}$ (~290 µV/K at 650 K)[11], $YbZn_{2-x}Mn_xSb_2$(~165 µV/K at 750 K)[25], $Yb_9Mn_{4.2}Sb_9$ (~190 µV/K at 950 K)[26], and $Yb_{14}MnSb_{11}$(~185 µV/K at 1275 K)[27]. Furthermore, $\alpha(T)$ of $Ca_{10}GdCdSb_9$ presents an even more interesting scenario whereby the value of 175.2 µV/K achieved at room temperature is significantly high and which increases to about 345 µV/K at 600 K. With this observation, it is plausible to infer that the density of states effective masses for both compounds are very high. This notion is further supported by the single parabolic band (SPB) model presented *vide infra.* It is noted that at 600 K, $\alpha(T)$ of both phases still shows a strong temperature dependence with no sign of the excitation of the minority carriers or features of mixed conduction. In $Yb_{10}LaCdSb_9$ and $Yb_{11}TrSb_9$ ($Tr$ = Ga, In)[13], features of mixed conduction sets in at elevated temperatures, and it is likely that the title phases too would manifest features of mixed conduction near 1000 K. Regardless, the additional features discussed below strongly supports the realization of excellent thermoelectric properties in these phases.

In Figure 2(b), the temperature dependence of electrical resistivity $\rho(T)$ of both phases are presented in the temperature range of 300 to 460 K. $\rho(T)$ of $Ca_{10}LaCdSb_9$ presented on the right axis decreases upon heating in a semiconducting manner. We note that this feature is in line with the observation in $Yb_{10}LaCdSb_9$[16], albeit the magnitude of $\rho(T)$ in $Ca_{10}LaCdSb_9$ (41.8 m$\Omega$·cm at 300 K) is about an order of magnitude higher. Although the origin of such a relatively higher value is not immediately apparent, the evolution of $\rho(T)$ in a semiconducting manner can result in significantly high $\rho(T)$ values. Conversely, $\rho(T)$ of $Ca_{10}GdCdSb_9$ shown on the left axis evolves in a metallic manner and features relatively low values that vary from about 2.94 to 3.5 m$\Omega$·cm at 300 and 460 K, respectively. The metallic nature of $\rho(T)$ and the linear temperature dependence of $\alpha(T)$ indicate a heavily doped, degenerate p-type semiconducting material that reminiscence several excellent thermoelectric materials. The thermoelectric transport properties of this phase, namely the enhance $\alpha(T)$ magnitude, relatively low $\rho(T)$ values and the inherent low thermal conductivity, would prove invaluable in achieving an excellent $zT$.

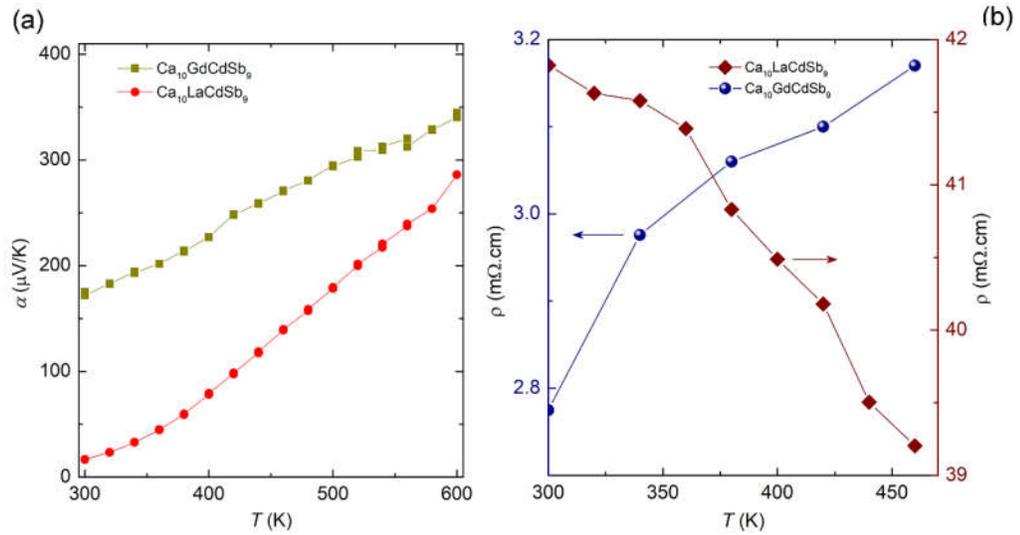

Figure 2. (a) Temperature dependence of Seebeck coefficient $\alpha(T)$. (b) Temperature dependence of electrical resistivity $\rho(T)$ of $Ca_{10}RECdSb_9$ ($RE$ = La, Gd).

Figure (3) presents results from Hall effect $R_H$ measurements on $Ca_{10}GdCdSb_9$ and are further analyzed using the single parabolic band (SPB) model. These results confirm a p-type material and indicate a room temperature value of Hall carrier concentration $n_H = 5.1 \times 10^{20}$ cm$^{-3}$, which is calculated from Hall coefficient: $R_H = 1/n_H e$, where $e$ is the electric charge. The Hall mobility $\mu_H$ on the other hand, is calculated from $R_H$ and $\rho$ as $\mu_H = R_H/\rho$. The temperature dependence of $n_H$ and $\mu_H$ are presented in Figure 3(a) and (b), respectively. $n_H$ steadily increases with temperature and attains a value $9.7 \times 10^{20}$ cm$^{-3}$ at 460 K. $\mu_H$ on the other hand, is observed to decrease with an increasing temperature with a relatively small room temperature value of ca. 5.0 cm$^2$/V·s, a feature which is a reminiscence of heavily doped semiconductors. A power law models the temperature dependence of $\mu_H$: $\mu_H \propto T^{-p}$, within a single scattering limit. The solid line in the plot is the fit to this expression with $p < 1.5$. The small $\mu_H$ observed can be associated with the heavy carrier mass near the Fermi level in Ca$_{10}$GdCdSb$_9$. We provide additional support to this as follows. $n_H$ is related to the chemical carrier concentration $n$ by:

$$n = r_H n_H, \tag{1}$$

where $r_H$ is the Hall factor and it is expressed as[28]:

$$r_H = \frac{3}{2} F_{1/2}(\eta) \frac{(1/2 + 2\lambda) F_{2\lambda - 1/2}(\eta)}{(1 + \lambda)^2 F_\lambda^2(\eta)}, \tag{2}$$

$\lambda$ relates to energy dependence of the carrier relaxation time $\tau$: $\tau = \tau_0 \epsilon^{\lambda - 1/2}$, which determines the scattering exponent. $F_j(\eta)$ is the Fermi integral defined as:

$$F_j(\eta) = \int_0^\infty \frac{\zeta^j d\zeta}{1 + e^{(\zeta - \eta)}}, \tag{3}$$

where $\eta$ and $\zeta$ are the reduced electrochemical potentials and reduced carrier energy, respectively. To determine $r_H$ from on the above set of equations, one would need to first determine the value of $\eta$ based on the analysis of the experimentally observed Seebeck coefficient $\alpha$ according to the equation:

$$\alpha = \pm \frac{k_B}{e} \left( \frac{(2 + \lambda) F_{(1+\lambda)}(\eta)}{(1 + \lambda) F_\lambda(\eta)} - \eta \right), \tag{4}$$

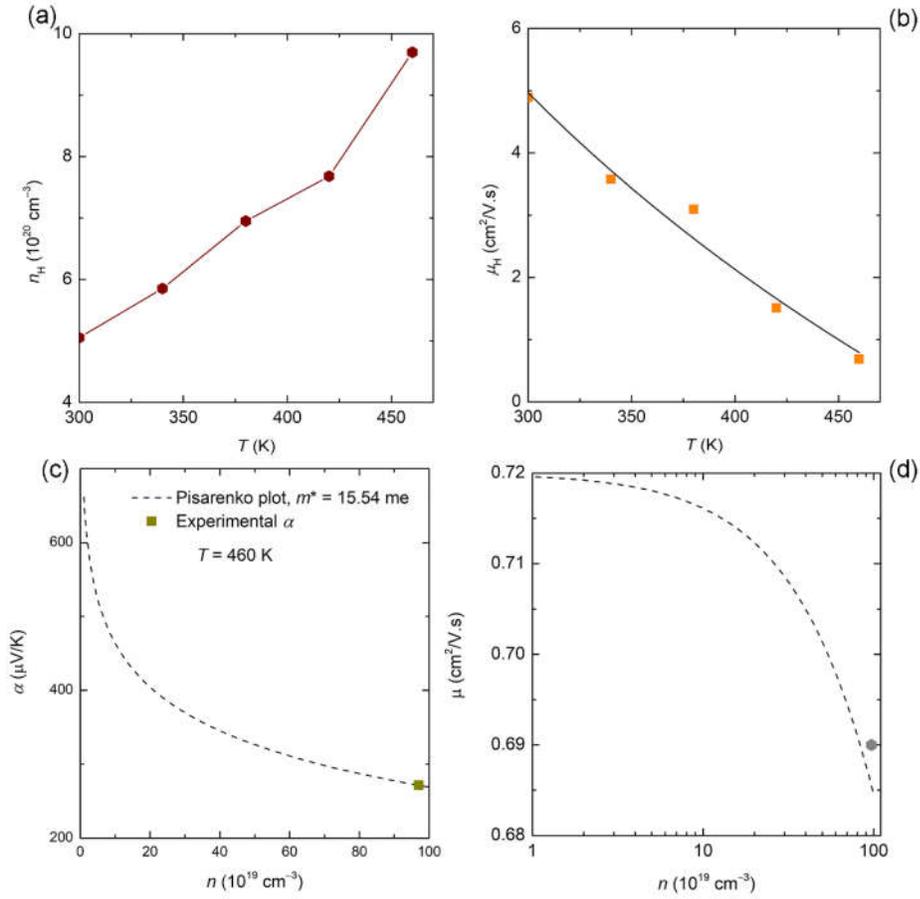

Figure 3. (a) The temperature dependence of Hall concentration and (b) Hall mobility. The solid line is a least-square fit to $\mu_H \propto T^{-p}$, with $p < 1.5$. (c) Carrier concentration variation of $\alpha(T)$ based on Pisarenko plot (d) Carrier concentration variation of mobility of $Ca_{10}GdCdSb_9$ using the SPB model.

where $k_B$ and $e$ are the Boltzmann constant and electric charge, respectively. Under the assumption that the charge carriers are limited by acoustic phonons ($\lambda = 0$), the thermoelectric properties obtained at room temperature are tabulated in Table 2. Here the magnitude of the calculated effective mass $m^*$ at room temperature is ca. $6.84m_e$, where $m_e$ is the rest mass of an electron, which is significantly high and is in support of the high $\alpha(T)$ observed at this temperature. It is also noted that $m^*$ increases with temperature and achieves a magnitude of $15.54m_e$ at 460 K. Such a feature can be associated with the temperature dependence of the bandgap. Also, the observed $r_H$ of 1.16 at 460 K is comparable to the value of 1.18 for acoustic phonons scattering mechanism[28].

In Figure 3(c), The Pisarenko plot of $\alpha$ against $n$ for $m^* = 15.54m_e$ at 460 K assuming an acoustic phonon scattering mechanism in SPB is presented based on the following expression[29]:

$$n = 4\pi \left( \frac{2m^* k_B T}{h^2} \right)^{3/2} \frac{F_{1/2}(\eta)}{r_H}, \tag{5}$$

where $h$ is the Planck constant. The Pisarenko relation is shown as a dashed line and the experimentally determined $\alpha$ is also shown. Also based on the same model, $\mu_H$ can be written as:

$$\mu_H = \mu_0 \frac{(1/2 + 2\lambda) F_{2\lambda - 1/2}(\eta)}{(1+\lambda) F_\lambda(\eta)}, \tag{6}$$

where $\mu_0$ relates to $\tau_0$: $\mu_0 = e\tau_0/m^*$. A plot of $\mu_H$ as a function of $n$ is shown in Figure 3(d).

**Table 2**: Selected transport properties of $Ca_{10}GdCdSb_9$ at 300 and 460 K.

| $T$ (K) | $\alpha$ (μV/K) | $\eta$ | $m^*$ ($m_e$) | $r_H$ |
|---|---|---|---|---|
| 300 | 175.2 | 0.51 | 6.84 | 1.12 |
| 460 | 271.3 | –0.99 | 15.54 | 1.16 |

The nature of the intricate bonding of the constituent atoms in this structure is known to mediate robust phonon scatterings, which can lead to low thermal conductivity $\kappa$. Also, since the title phase is isostructural to $Yb_{10}LaCdSb_9$, with similar structural bonding and unit cell volume, their $\kappa$ are expected to be comparable. Figure S4(a) presents an extrapolated $\kappa_{Tot}(T)$ data from $Yb_{10}LaCdSb_9$[16] in the temperature range of 300 to 460 K. The data feature low values for the total thermal conductivity $\kappa_{Tot}$ which varies from the room temperature value of ca. 0.90 to 0.84 W/m·K at 460 K. Also shown in the same panel is the lattice thermal conductivity $\kappa_{Latt}$ which is deconfined from $\kappa_{Tot}$ as $\kappa_{Latt} = \kappa_{Tot} - \kappa_{Elec}$. Where $\kappa_{Elec}$ is the electronic contribution to the thermal conductivity, which is given by the Wiedemann-Franz relation[30]:

$$\kappa_{Elec} = \frac{L_0 T}{\rho}, \tag{7}$$

where $L_0$ is the Lorentz number given by $\pi^2 k_B^2/3e^2 = 2.45 \times 10^{-8}$ W$\Omega$/K$^2$ and $\rho$ is the electrical resistivity of Ca$_{10}$GdCdSb$_9$ shown in Figure 2(b). The estimated $\kappa_{Latt}$ varies between ca. 0.71 and 0.62 W/m·K at 300 to 460 K, respectively. While the estimation of $\kappa_{Latt}$ by the above method may be valid at room temperature, it is unsuitable for certain metallic and degenerate semiconductors where the chemical potential is strongly temperature dependent. Also, the Lorentz number $L$ is directly proportional to $\eta$ and for several degenerate semiconductors, they are found to be much smaller than $L_0$ at high temperatures. The Lorentz number $L$ as a function of $\eta$ using the SPB model is given as:

$$L = \frac{k_B^2}{e^2} \frac{(1+\lambda)(3+\lambda) F_\lambda(\eta) F_{\lambda+2}(\eta) - (2+\lambda)^2 F_{\lambda+1}(\lambda)^2}{(1+\lambda)^2 F_\lambda(\eta)}. \tag{8}$$

For a carrier mobility limited by acoustic scattering ($\lambda = 0$) and the value of $\eta$ from the analysis of $\alpha(T)$, the obtained values of $L$ are presented in Figure S4(b). $L$ decreases with increasing temperature, with values that are much lower than $L_0$ in line with findings in several thermoelectric materials[28,29].

**Thermoelectric efficiency**

The temperature dependence of thermoelectric power factor $PF$ ($=\alpha^2/\rho$) and dimensionless thermoelectric figure of merit $zT$ calculated from the experimental results of $\alpha(T)$, $\rho(T)$ and $\kappa(T)$ are presented in Figures 4(a) and S5, respectively. The determined $PF$ ranges from 11.1 µW/cm·K$^2$ at 300 K to 23.2 µW/cm·K$^2$ at 460 K. To put the magnitude of the observed $PF$ into perspectives, a plot of $PF$ of the title phase Ca$_{10}$GdCdSb$_9$ and those of selected Zintl phases such as Ca$_{10}$InSb$_9$[9], Yb$_{10}$LaCdSb$_9$[16], Yb$_{14}$MnSb$_{11}$[27], YbCd$_{2-x}$Zn$_x$Sb$_2$ ($x = 0.4$)[31], and EuZn$_{2-x}$Cd$_x$Sb$_2$ ($x$

= 0.1)$^{32}$ is presented in Figure 4(b). The magnitude of PF of the title phase at 460 K is higher than those of the other phases and with features that suggest a further increase in its value with increasing temperature. These observations along with the putative low thermal conductivity in this phase present an interesting scenario where intricate atomic bonding and a large unit-cell volume conspires with a narrow bandgap in a material to produce remarkable thermoelectric properties. At room temperature, Ca$_{10}$GdCdSb$_9$ achieves a zT of 0.47 and reaches a remarkably high value of 1.73 at 460 K (see Figure S5). Such value is comparable to that of state-of-the-art material PbTe$^{33}$ with zT > 1.5 at 773 K and much enhanced than Yb$_{14}$MnSb$_{11}$$^{27}$ with a zT ~ 1 at 1223 K. Although the estimation of zT is limited to 300–460 K, the temperature dependence of the various thermoelectric parameters is in favor of a much-enhanced value for T > 460 K. In addition, a theoretical zT and PF can be calculated based on the SPB model. For such theoretical calculations, the relevant parameters are defined as:

$$zT = \frac{\alpha^2}{L + (\psi\beta)^{-1}}, \tag{9}$$

where

$$\psi = \frac{8\pi e}{3}\left(\frac{2m_e k_B}{h^2}\right)^{3/2}(1+\lambda)F_\lambda(\eta), \tag{10}$$

and

$$\beta = \frac{\mu_0\left(m^*/m_e\right)^{3/2}T^{5/2}}{\kappa_{Latt}}. \tag{11}$$

The calculated zT based on the SPB model in the temperature range of 300 to 460 K is presented in Figure S6. Here zT values varies from the room temperature value of ca.

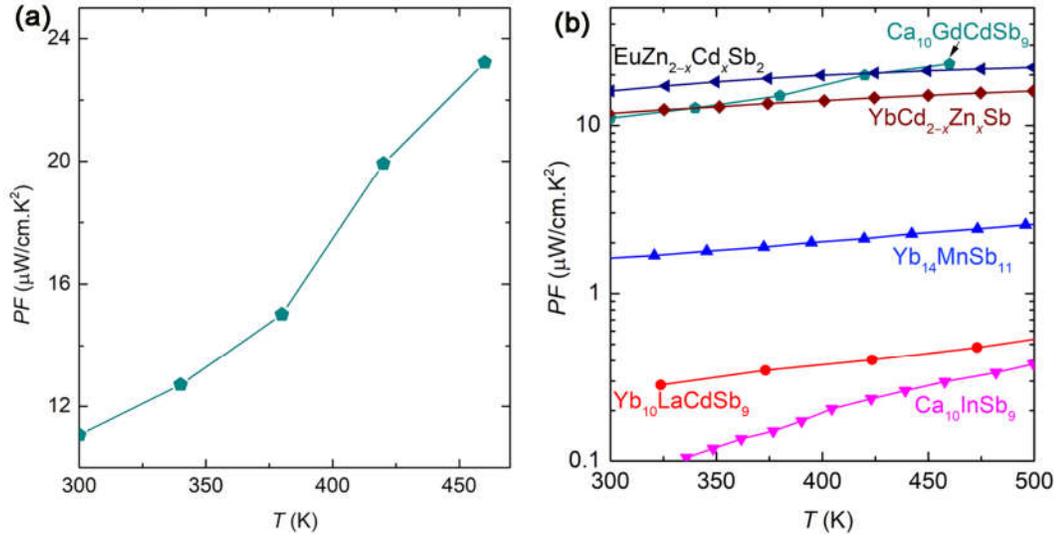

Figure 4. (a) The temperature dependence thermoelectric power factor *PF* of $Ca_{10}GdCdSb_9$, (b) *PF* of $Ca_{10}GdCdSb_9$ together with those of other Zintl phases. The y-axis is presented on a log scale for clarity.

0.41 to a peak value of ca. 0.7 at 380 K. Here, the calculated values clearly underestimate the experimentally observed *zT*. The observed deviation may likely be an indication of the complex band structure in $Ca_{10}GdCdSb_9$ which may be associated with the presence of multiple degenerate bands close to the conduction band edge. In such an instance, the determination of $m^*$, *L* and $\kappa_{Latt}$ might be prone to errors and consequently impact on the calculated *zT* and optimal $n$[34]. Although the theoretical calculation using the SPB model seems to underestimate *zT* and *PF*, such calculation still provide support for the excellent thermoelectric properties in the title compounds and relevant insight into the tuning of n near the Fermi energy.

**Conclusions**

This report presents two highly disordered and complex Zintl phases $Ca_{10}RECdSb_9$ (*RE* = La, Gd) with the orthorhombic *Ibam* space group, which can be harnessed and developed into an efficient material for use in thermoelectric energy generation. These compounds feature large α (286.2 µV/K and 345 µV/K at 600 K for La and Gd phases, respectively) driven by a large density of states effective mass near the Fermi level. $\rho(T)$ of $Ca_{10}LaCdSb_9$ is semiconducting with relatively large magnitudes, while $Ca_{10}GdCdSb_9$ is reminiscent of heavily doped semiconductors with low $\rho(T)$ values that translate into large electrical conductivity. In addition, these phases are known to feature low intrinsic $\kappa_{Latt}$ (~ 0.71 W/m·K), which is attributed to factors such as their complex atomic bonding and large unit cell volumes which determine the number of phonons states within the first Brillouin zone[35]. The experimentally determined $PF$ of 23.2 µW/cm·$K^2$ at 460 K is remarkably high.

Theoretical calculations employing the SPB model were further carried out to provide additional perspective into the thermoelectric properties of $Ca_{10}GdCdSb_9$. These calculations gave support to the high density of state effective mass as well as the realization of high $zT$ value. Albeit the calculated $zT$ and $PF$ lags that of the experimentally determined values. We posit that these discrepancies lie in the complex nature of the band structure with multiple degenerate bands close to the conduction band maximum which can result in a non-precise determination of the various transport parameters. Future studies on these materials will benefit from the determination of their electronic band structure, which will require high computational cost considering the nature of the disorder as well the large unit cell volume.

## Conflicts of interest

There are no conflicts to declare.

## Acknowledgements

This work was supported by the U.S. Department of Energy, Office of Science, Basic Energy Sciences through grant DE-SC0008885.

# SUPPORTING INFORMATION

The Highly Disordered Zintl Phase $Ca_{10}GdCdSb_9$ – New Example of a p-type Semiconductor with Remarkable Thermoelectric Properties

*Michael O. Ogunbunmi and Svilen Bobev[*]*

Department of Chemistry and Biochemistry, University of Delaware, Newark, Delaware, 19716, United States

[*]Corresponding author; Email: bobev@udel.edu

**Table of Contents**

**Table S1**: Fractional Atomic Coordinates and Equivalent Isotropic Displacement Parameters ($U_{eq}$ /Å$^2$) Ca$_{10}$GdCdSb$_9$. $U_{eq}$ is defined as 1/3 of the trace of the orthogonalized $U_{ij}$ tensor.

| Atom | Wyckoff | x | y | z | $U_{eq}$ |
|---|---|---|---|---|---|
| Sb1[a] | 16k | 0.15483(14) | 0.1460(2) | 0.31139(8) | 0.0132(4) |
| Sb2[b] | 16k | 0.17147(18) | 0.1776(3) | 0.31877(10) | 0.0138(4) |
| Sb3[c] | 8j | 0.03611(4) | 0.39128(4) | 0 | 0.0123(2) |
| Sb4 | 8j | 0.13676(4) | 0.11946(5) | 0 | 0.0168(2) |
| Sb5[d] | 8j | 0.3728(11) | 0.0295(11) | 0 | 0.023(4) |
| Sb6 | 4b | 1/2 | 0 | 1/4 | 0.0108(2) |
| Cd1[e] | 8h | 0 | 0 | 0.11371(7) | 0.0176(2) |
| Cd2[f] | 4a | 0 | 0 | 1/4 | 0.014(10) |
| Ca1/Gd1[g] | 16k | 0.08129(7) | 0.27333(7) | 0.15809(5) | 0.0145(3) |
| Ca2[h] | 16k | 0.2767(18) | 0.0695(16) | 0.1676(14) | 0.009(6) |
| Ca3/Gd2[i] | 16k | 0.31490(9) | 0.05743(9) | 0.12939(8) | 0.0166(4) |
| Ca4/Gd3[j] | 8j | 0.33799(12) | 0.32707(12) | 0 | 0.0179(5) |
| Ca5[k] | 8h | 0 | 0 | 0.1747(3) | 0.016(1) |

Site occupancies: [a]Sb1 = 0.598(9); [b]Sb2 = 0.402(8); [c]Sb3 = 0.959(3); [d]Sb5 = 0.044(3); [e]Cd1 = 0.6128; [f]Cd2 = 0.025(4); [g]Ca1/Gd1 = 0.901(2)Ca1 + 0.099Gd1; [h]Ca2 = 0.053(6); [i]Ca3/Gd2 = 0.972(2)Ca3 + 0.028(2)Gd2; [j]Ca4/Gd3 = 0.959(3)Ca4 + 0.041(3)Gd3; [k]Ca5 = 0.403(7)

**Table S2**. Selected crystallographic data of $Ca_{10}LaCdSb_9$ measured at 200(2) K, space group *Ibam*, $Z = 4$, Mo Kα radiation $\lambda = 0.71073$ Å.

| | |
|---|---|
| refined formula | $Ca_{9.93(1)}Cd_{1.31(1)}La_{0.78}Sb_{9.01(1)}$ |
| formula weight | 1751.12 |
| crystal system | orthorhombic |
| $a$/Å | 11.873(6) |
| $b$/Å | 12.472(6) |
| $c$/Å | 16.872(8) |
| $V$/Å$^3$ | 2499(2) |
| $\rho_{calc}$/ g cm$^{-3}$ | 4.66 |
| $\mu$/cm$^{-1}$ | 139.9 |
| $R_1$ $(I \geq 2\sigma_I)$[a] | 0.0310 |
| $wR_2$ $(I \geq 2\sigma_I)$[a] | 0.0471 |
| $R_1$ (all data) | 0.0401 |
| $wR_2$ (all data) | 0.0488 |
| Largest difference peak and hole | 1.23; −0.91 |

$R_1 = \sum ||F_o| - |F_c||/\sum |F_o|$; $wR_2 = [\sum[w(F_o^2 - F_c^2)^2]/\sum[w(F_o^2)^2]]^{1/2}$, where $w = 1/[\sigma^2 F_o^2 + (0.0093P)^2 + (22.2781P)]$, and $P = (F_o^2 + 2F_c^2)/3$.

**Table S3**: Fractional Atomic Coordinates and Equivalent Isotropic Displacement Parameters (Å$^2$) Ca$_{10}$LaCdSb$_9$. $U_{eq}$ is defined as 1/3 of the trace of the orthogonalized $U_{ij}$ tensor.

| Atom | Wyckoff | x | y | z | $U_{eq}$ |
|---|---|---|---|---|---|
| Sb1[a] | 16k | 0.1554(3) | 0.1466(4) | 0.31155(16) | 0.0144(5) |
| Sb2[b] | 16k | 0.1718(4) | 0.1756(7) | 0.3185(2) | 0.0158(6) |
| Sb3[c] | 8j | 0.03608(4) | 0.39123(4) | 0 | 0.01502(14) |
| Sb4 | 8j | 0.13652(4) | 0.11885(4) | 0 | 0.01792(13) |
| Sb5[d] | 8j | 0.3750(19) | 0.0270(30) | 0 | 0.032(10) |
| Sb6 | 4b | 1/2 | 0 | 1/4 | 0.0110(14) |
| Cd1[e] | 8h | 0 | 0 | 0.11583(5) | 0.0222(2) |
| Cd2[f] | 4a | 1/2 | 1/2 | 3/4 | 0.022(10) |
| Ca1 | 16k | 0.08186(8) | 0.27380(8) | 0.15879(5) | 0.0162(2) |
| Ca2/La1[g] | 16k | 0.31343(6) | 0.05852(7) | 0.13091(5) | 0.0198(3) |
| Ca3/La2[h] | 8j | 0.33885(9) | 0.32802(8) | 0 | 0.0178(3) |
| Ca4[i] | 8h | 0 | 0 | 0.1749(3) | 0.0222(2) |

Site occupancies: [a]Sb1 = 0.617(18); [b]Sb2 = 0.385(18); [c]Sb3 = 0.981(2); [d]Sb5 = 0.020(2); [e]Cd1 = 0.6453(19); [f]Cd2 = 0.024(3); [g]Ca2/La1 = 0.883(2)Ca2 + 0.117(2)La1; [h]Ca3/La2 = 0.844(3)Ca3 + 0.156(3)La2; [i]Ca4 = 0.3547(19).

**Table S4.** Selected distances (Å) for $Ca_{10}GdCdSb_9$.

| Atom pairs | | Distance/Å |
| --- | --- | --- |
| Sb6 | Ca1 | 3.3521(9)×3 |
| Sb6 | Ca3/ Gd2 | 3.064(1) ×4 |
| Sb4 | Ca3/ Gd2 | 3.120(1) ×2 |
| Sb6 | Ca2 | 3.10(2) ×4 |
| Sb3 | Sb3 | 2.835(1) |
| Sb3 | Ca1/ Gd1 | 3.0746(9) ×2 |
| Sb4 | Cd1 | 2.9073(8) ×2 |
| Sb4 | Ca1 | 3.3327(9) ×2 |
| Sb4 | Sb5 | 3.01(1) |
| Sb1 | Cd1 | 2.868(2) |
| Sb1 | Ca1 | 3.251(1) |
| Sb1 | Ca1/ Gd1 | 3.140(1) |
| Sb1 | Ca1 | 3.318(3) |
| Sb1 | Ca4 | 3.179(2) |
| Sb1 | Ca3 | 3.313(2) |
| Sb3 | Ca3/ Gd2 | 3.475(1) ×2 |
| Sb3 | Ca3/ Gd2 | 3.457(1) ×2 |
| Cd1 | Sb2 | 3.204(3) ×2 |
| Cd2 | Sb2 | 3.212(4) ×4 |
| Cd1 | Ca2 | 3.50(2) ×2 |
| Sb2 | Ca4/Gd3 | 3.039(2) ×2 |
| Sb5 | Sb5 | 3.10(3) |
| Sb5 | Ca2 | 3.07(3) ×2 |
| Ca5 | Sb2 | 3.000(4) ×2 |
| Sb4 | Ca4 | 3.511(2) |
| Sb4 | Ca4 | 3.596(2) |
| Sb4 | Ca2 | 3.32(2) ×2 |

Table S5. Selected distances (Å) for $Ca_{10}LaCdSb_9$.

| Atom pairs | | Distance/Å |
| --- | --- | --- |
| Sb6 | Ca1 | 3.357(1) ×4 |
| Sb6 | Ca2/La2 | 3.078(1) ×4 |
| Sb3 | Sb3 | 2.845(1) |
| Sb3 | Ca1 | 3.101(1) ×2 |
| Sb3 | Ca3 | 3.680(2) |
| Sb3 | Ca3/La3 | 3.600(2) |
| Sb3 | Ca2 | 3.501(1) ×2 |
| Sb3 | Ca2/La2 | 3.525(1) ×2 |
| Sb4 | Cd1 | 2.940(1) ×2 |
| Sb4 | Ca1 | 3.366(2) ×4 |
| Sb4 | Ca3 | 3.596(2) |
| Sb4 | Ca3 | 3.546(2) |
| Sb4 | Ca2 | 3.139(1) ×2 |
| Sb4 | Sb5 | 3.06(2) |
| Sb1 | Ca3 | 3.196(3) |
| Sb1 | Ca2 | 3.318(3) |
| Sb1 | Ca2 | 3.744(3) |
| Sb1 | Ca4 | 3.473(7) |
| Cd1 | Ca3 | 3.479(1) ×2 |
| Cd1 | Ca4 | 3.531(5) |
| Cd1 | Sb2 | 3.192(8) ×4 |
| Ca1 | Sb2 | 3.016(7) |
| Ca1 | Sb2 | 3.274(3) |
| Ca1 | Sb2 | 3.146(3) |
| Ca3 | Sb2 | 3.066(4) ×2 |
| Ca3 | Sb5 | 3.55(3) |
| Ca2/La2 | Sb2 | 3.428(9) |
| Ca2 | Sb2 | 3.477(5) |
| Ca4 | Sb2 | 2.996(9) ×2 |
| Sb2 | Cd2 | 3.209(9) |
| Sb5 | Sb5 | 3.04(5) |

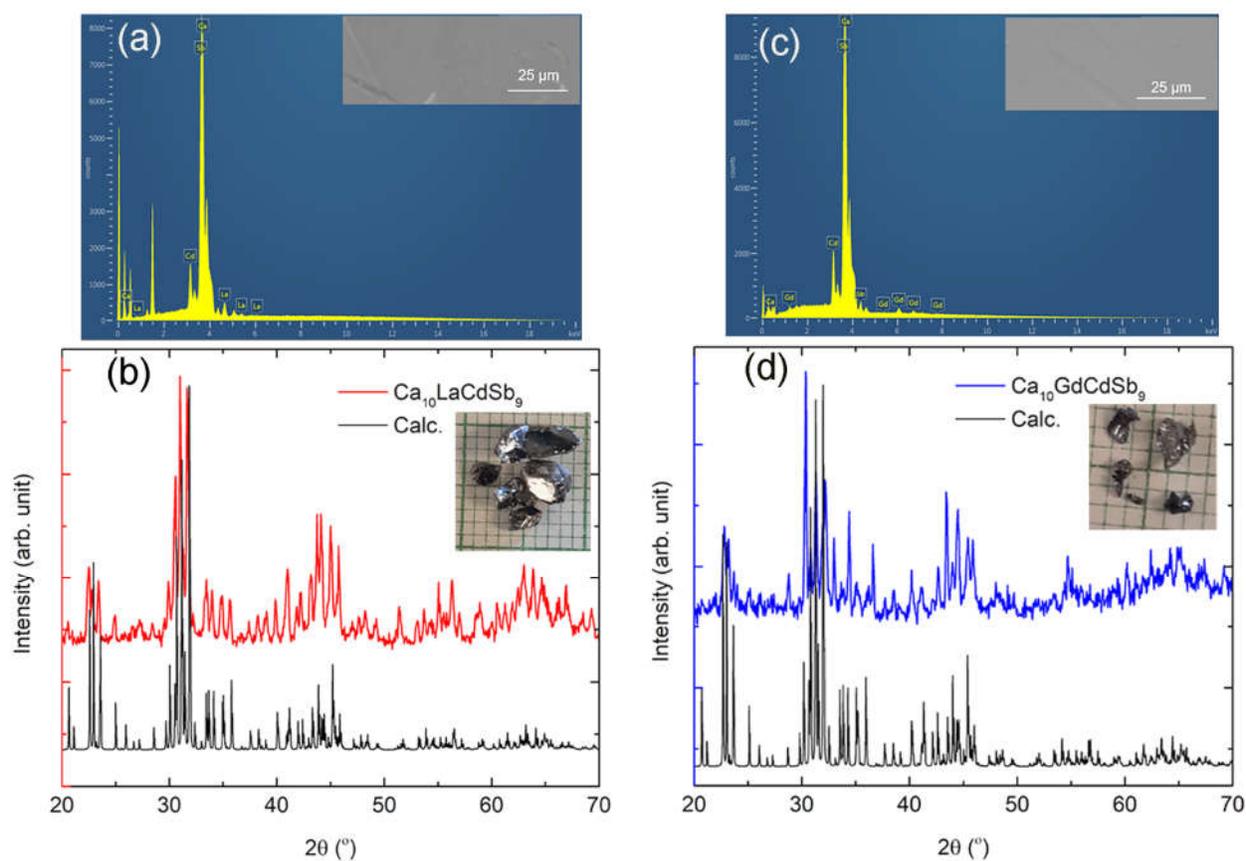

Figure S1. (a) Representative histogram from the EDS analysis of $Ca_{10}LaCdSb_9$ under high magnification SEM image shown in the inset with compositions comparable to that refined from single crystal data. (b) Powder XRD (PXRD) pattern of $Ca_{10}LaCdSb_9$ together with the calculated pattern based on single crystal XRD structural elucidation. The inset are the single crystals of $Ca_{10}LaCdSb_9$ (c) The EDS analysis of $Ca_{10}GdCdSb_9$ with the SEM image shown as inset. (d) PXRD pattern of $Ca_{10}GdCdSb_9$ together with the calculated pattern. Single crystals of $Ca_{10}GdCdSb_9$ are shown in the inset.

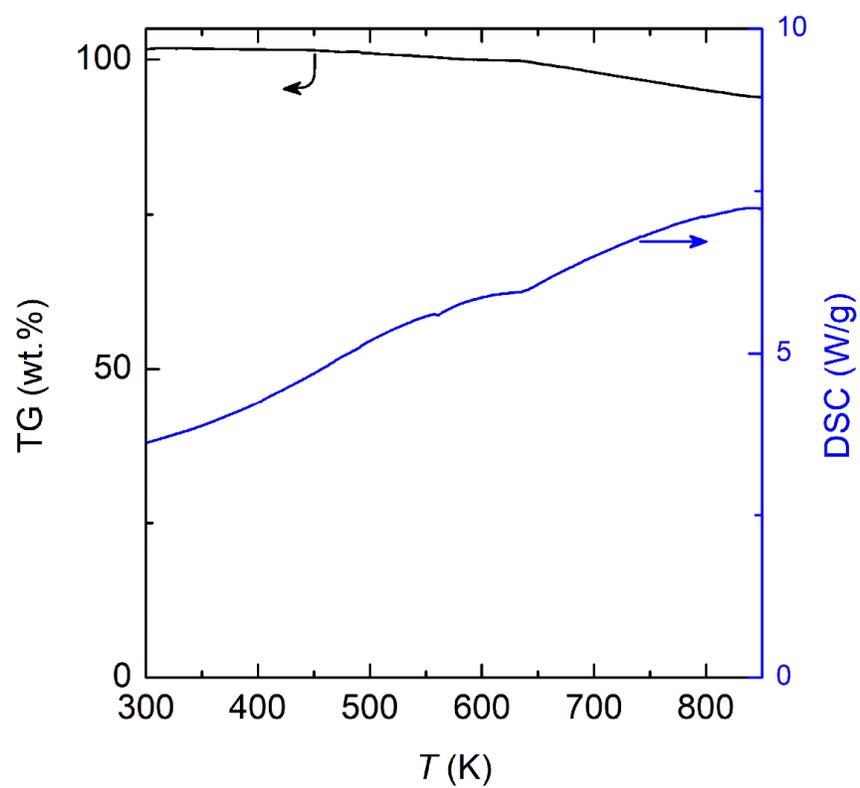

Figure S2. Plots of TG/DSC of $Ca_{10}GdCdSb_9$. The weight (wt.%) is represented by a black line while the heat flow (W/g) is represented by a blue line.

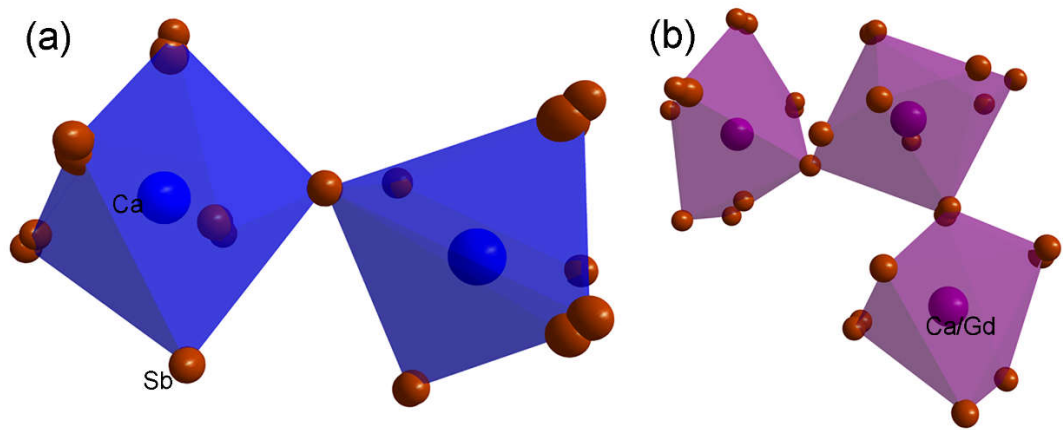

Figure S3. (a) and (b) Coordination environment of Ca and Ca/Gd sites of Ca$_{10}$GdCdSb$_9$, respectively.

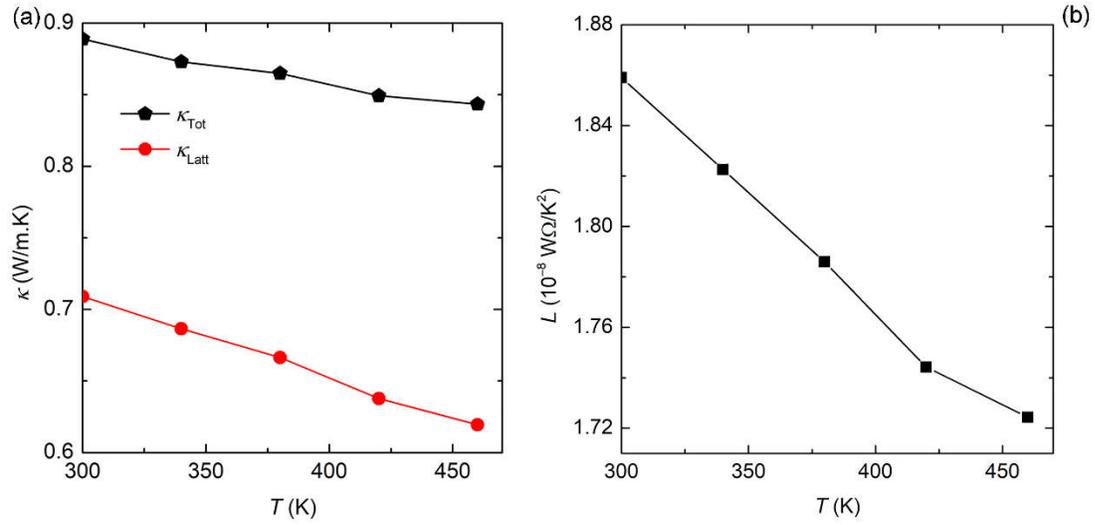

Figure S4. (a) Temperature dependence of total and lattice thermal conductivity $\kappa(T)$ extrapolated from $Yb_{10}LaCdSb_9$[16]. (b) Temperature variation of Lorentz number $L(T)$.

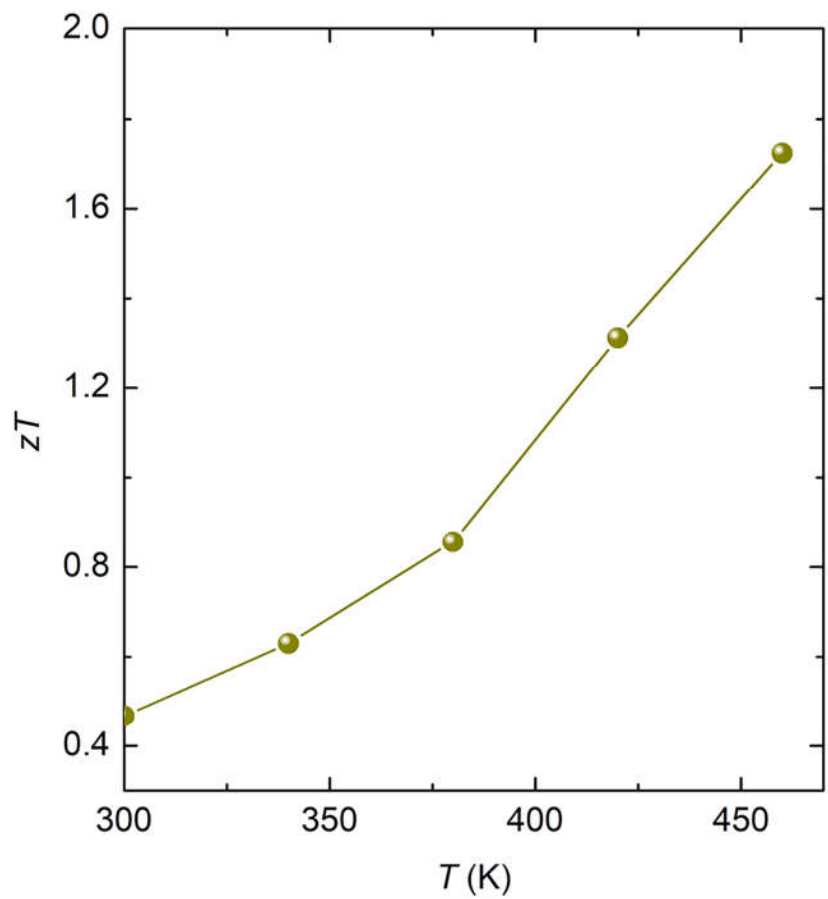

Figure S5. Temperature dependence of dimensionless thermoelectric figure of merit $zT$ of $Ca_{10}GdCdSb_9$ based on experimental data.

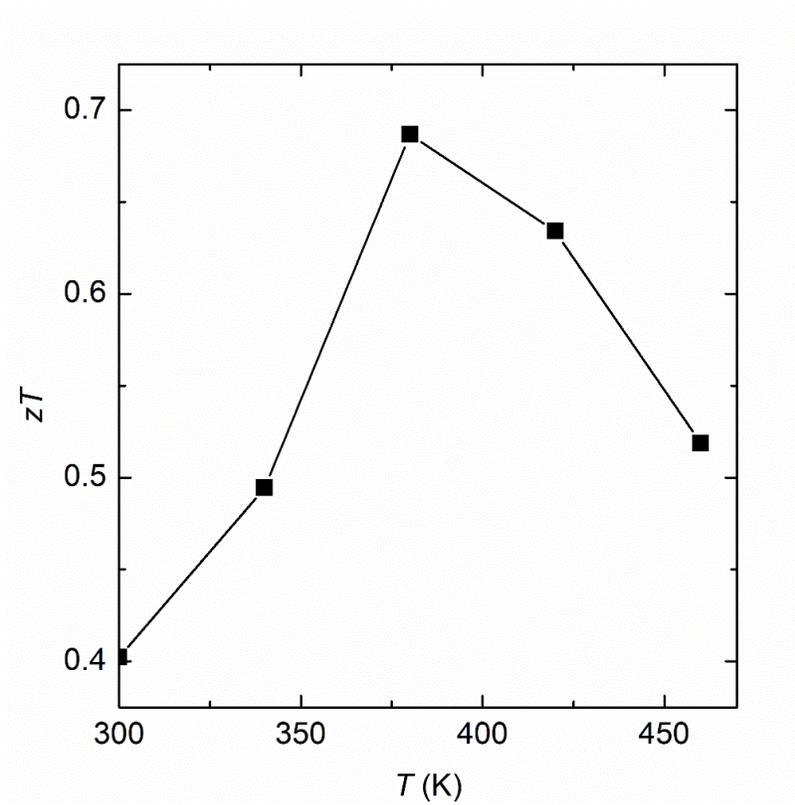

Figure S6. Temperature dependence of $zT$ of $Ca_{10}GdCdSb_9$ based on the SPB model.